\documentclass[12pt]{article}
\usepackage{amssymb,graphicx} 

\begin{document}

\newcommand{\sect}[1]{\setcounter{equation}{0}\section{#1}}
\renewcommand{\theequation}{\thesection.\arabic{equation}}
\newcommand{\be}{\begin{equation}}
\newcommand{\ee}{\end{equation}}
\newcommand{\bea}{\begin{eqnarray}}
\newcommand{\eea}{\end{eqnarray}}
%%%%%%%%%%%%%%%%%%%%%%%%%
%%%%%%%%     GREQUES     %%%%%%%
%%%%%%%%%%%%%%%%%%%%%%%%%
\newcommand{\eps}{\epsilon}
\newcommand{\om}{\omega}
\newcommand{\vph}{\varphi}
\newcommand{\sig}{\sigma}
%%%%%%%%%%%%%%%%%%%%%%%%
%%%%%%%  C, R, Q, Z, N, Id  %%%%%
%%%%%%%%%%%%%%%%%%%%%%%%
%---- CORPS DES COMPLEXES
\newcommand{\CC}{\mbox{${\mathbb C}$}}
%---- CORPS DES REELS
\newcommand{\RR}{\mbox{${\mathbb R}$}}
%---- CORPS DES RATIONNELS
\newcommand{\QQ}{\mbox{${\mathbb Q}$}}
%---- GROUPE DES ENTIERS
\newcommand{\ZZ}{\mbox{${\mathbb Z}$}}
%---- NATURELS
\newcommand{\NN}{\mbox{${\mathbb N}$}}
%---- IDENTITE EN 12 PT
\newcommand{\1}{\mbox{\hspace{.0em}1\hspace{-.24em}I}}
\newcommand{\II}{\mbox{${\mathbb I}$}}
%%%%%%%%%%%%%%%%%%%%%%%%%%%%%%%%%
%%%%%%%%    DIVERS   %%%%%%%%%%%%
%%%%%%%%%%%%%%%%%%%%%%%%%%%%%%%%%
\newcommand{\prt}{\partial}
\newcommand{\und}[1]{\underline{#1}}
\newcommand{\wh}[1]{\widehat{#1}}
\newcommand{\wt}[1]{\widetilde{#1}}
\newcommand{\mb}[1]{\ \mbox{\ #1\ }\ }
\newcommand{\half}{\frac{1}{2}}
\newcommand{\noin}{\not\!\in}
\newcommand{\rhotimes}{\mbox{\raisebox{-1.2ex}{$\stackrel{\displaystyle\otimes}
{\mbox{\scriptsize{$\rho$}}}$}}}
\newcommand{\bin}[2]{{\left( {#1 \atop #2} \right)}}
\newcommand{\ri}{{\rm i}}
\newcommand{\rd}{{\rm d}}
\newcommand{\A}{{\cal A}}
\newcommand{\B}{{\cal B}}
\newcommand{\C}{{\cal C}}
\newcommand{\F}{{\cal F}}
\newcommand{\E}{{\cal E}}
\newcommand{\cP}{{\cal P}}
\newcommand{\R}{{\cal R}}
\newcommand{\T}{{\cal T}}
\newcommand{\W}{{\cal W}}
\newcommand{\cS}{{\cal S}}
\newcommand{\bS}{{\bf S}}
\newcommand{\cL}{{\cal L}}
\newcommand{\cV}{{\cal V}}
\newcommand{\hlp}{{\RR}_+}
\newcommand{\hlm}{{\RR}_-}
\newcommand{\Hil}{{\cal H}}
\newcommand{\D}{{\cal D}}
\newcommand{\G}{{\cal G}}
\newcommand{\alg}{\C} 
\newcommand{\rep}{\F(\A)}
\newcommand{\trep}{\G_\beta(\A)}
\newcommand{\form}{\langle \, \cdot \, , \, \cdot \, \rangle }
\newcommand{\e}{{\rm e}}
\newcommand{\by}{{\bf y}}
\newcommand{\bp}{{\bf p}}
\newcommand{\LL}{\mbox{${\mathbb L}$}}
\newcommand{\Rp}{{R^+_{\, \, \, \, }}}
\newcommand{\Rm}{{R^-_{\, \, \, \, }}}
\newcommand{\Rpm}{{R^\pm_{\, \, \, \, }}}
\newcommand{\Tp}{{T^+_{\, \, \, \, }}}
\newcommand{\Tm}{{T^-_{\, \, \, \, }}}
\newcommand{\Tpm}{{T^\pm_{\, \, \, \, }}}
\newcommand{\baral}{\bar{\alpha}}
\newcommand{\barbt}{\bar{\beta}}
\newcommand{\supp}{{\rm supp}\, }
\newcommand{\Pt}{\widetilde{P}}
\newcommand{\At}{\widetilde{A}}
\newcommand{\Bt}{\widetilde{B}}
\newcommand{\St}{\widetilde{S}}
\newcommand{\jt}{\widetilde{j}}
\newcommand{\Jt}{\widetilde{J}}
\newcommand{\Gt}{\widetilde{G}}
\newcommand{\EE}{\mbox{${\mathbb E}$}}
\newcommand{\JJ}{\mbox{${\mathbb J}$}}
\newcommand{\MM}{\mbox{${\mathbb M}$}}
\newcommand{\ct}{{\cal T}}
\newcommand{\ph}{\varphi}
\newcommand{\phd}{\widetilde{\varphi}}
\newcommand{\phl}{\varphi_{{}_L}}
\newcommand{\phr}{\varphi_{{}_R}}
\newcommand{\phpl}{\varphi_{{}_{+L}}}
\newcommand{\phpr}{\varphi_{{}_{+R}}}
\newcommand{\phml}{\varphi_{{}_{-L}}}
\newcommand{\phmr}{\varphi_{{}_{-R}}}
\newcommand{\phpml}{\varphi_{{}_{\pm L}}}
\newcommand{\phpmr}{\varphi_{{}_{\pm R}}}
\newcommand{\Ei}{\rm Ei}

%%%%%%%% 
\newcommand{\finprf}{\null \hfill {\rule{5pt}{5pt}}\\[2.1ex]\indent}

%%%%%%%%%%%%%%%%%%%%%%%
\pagestyle{empty}
\rightline{January 2008}

\vfill

\begin{center}
{\Large\bf Quantum Field Theory on Star Graphs}
\\[2.1em]

\bigskip

{\large
B. Bellazzini$^{a}$\footnote{b.bellazzini@sns.it}, 
M. Burrello$^{a}$\footnote{m.burrello@sns.it} 
M. Mintchev$^{b}$\footnote{mintchev@df.unipi.it} 
and P. Sorba$^{c}$\footnote{sorba@lapp.in2p3.fr}}\\

\null

\noindent 

{\it 
$^a$ INFN and Scuola Normale Superiore, Piazza dei Cavalieri 7, 
56127 Pisa,  Italy\\[2.1ex]
$^b$ INFN and Dipartimento di Fisica, Universit\`a di
Pisa, Largo Pontecorvo 3, 56127 Pisa, Italy\\[2.1ex] 
$^c$ LAPTH, 9, Chemin de Bellevue, BP 110, F-74941 Annecy-le-Vieux 
cedex, France}
\vfill

\end{center}
\begin{abstract}

We discuss some basic aspects of quantum fields on star graphs, focusing 
on boundary conditions, symmetries and scale invariance in particular. 
We investigate the four-fermion bulk interaction in detail.  
Using bosonization and vertex operators, we solve the model exactly for 
scale invariant boundary conditions, formulated in terms of the fermion current 
and without dissipation. The critical points are classified and determined explicitly. 
These results are applied for deriving the charge and spin transport, which have  
interesting physical features. 

\end{abstract}
\bigskip 
\medskip 
\bigskip 

Talk presented at the Isaac Newton Institute programme {\it Analysis on 
Graphs and its Applications} (Cambridge 2007) -- to be published in Proceedings of 
Symposia in Pure Mathematics (AMS). 

\vfill
\rightline{IFUP-TH 1/2008}
\rightline{LAPTH-Conf-1234/08}
\newpage
\pagestyle{plain}
\setcounter{page}{1}

%%%%%%%%%%%%%%%%%%%%%%%%%%%%%%%%
\section{Introduction} 
\bigskip
The physics of quantum wires is nowadays a widely investigated topic, 
the main driving force being the fast development of nano-scale technology.  
Quantum wires are networks of junctions and wires which are very thin -- of 
the order of few nanometers. For this reason such devices, which 
are now created and tested in laboratory, are fairly well modeled by graphs. 
Among others, the most interesting physical problems concern the charge 
and the spin transport. The fact that these problems can be investigated in depth 
by means of quantum field theory (QFT) is our basic motivation for the construction and the study 
of quantum fields on graphs. In this context the development of bosonization and 
vertex algebras is essential and interesting also from the mathematical point of view. 

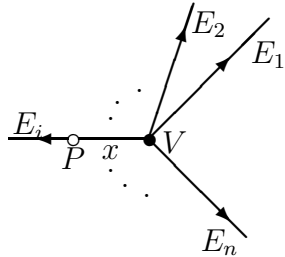
\begin{figure}[tb]
%\blankbox{.6\columnwidth}{5pc}
\setlength{\unitlength}{1mm}
\begin{picture}(20,20)(-42,20)
%names
\put(25.2,0.7){\makebox(20,20)[t]{$\bullet$}}
\put(28.5,1){\makebox(20,20)[t]{$V$}}
\put(42,11){\makebox(18,22)[t]{$E_1$}}
\put(33,17){\makebox(20,20)[t]{$E_2$}}
\put(9,3.5){\makebox(20,20)[t]{$E_i$}}
\put(15,-1.2){\makebox(20,20)[t]{$P$}}
\put(20,-0.8){\makebox(20,20)[t]{$x$}}
\put(34.5,-12){\makebox(20,20)[t]{$E_n$}}
%lines
\thicklines 
\put(35,20){\line(1,1){16}}
\put(35,20){\line(-1,0){9.3}}
\put(24.4,20){\line(-1,0){8}}
\put(35,20){\line(1,-1){13}}
\put(35,20){\line(1,3){6}}
%points
\put(20,3){\makebox(20,20)[t]{$.$}}
\put(20.9,5){\makebox(20,20)[t]{$.$}}
\put(23.8,6.6){\makebox(20,20)[t]{$.$}}
\put(20,-4){\makebox(20,20)[t]{$.$}}
\put(21.9,-6){\makebox(20,20)[t]{$.$}}
\put(24.8,-7.3){\makebox(20,20)[t]{$.$}}
\put(46,31){\vector(1,1){0}}
\put(46,9){\vector(1,-1){0}}
\put(40,35){\vector(1,3){0}}
\put(20,20){\vector(-1,0){0}}
\put(15,0.7) {\makebox(20,20)[t]{$\circ$}}
\end{picture} 
\vskip 1,5truecm
\caption{ A star graph $\Gamma$ with $n$ edges.} 
\label{stargraph}
\end{figure}

For simplicity, we consider below graphs of the type shown in Fig.~\ref{stargraph}.
They are called star graphs\footnote{For $n=3,4$ junctions of this type 
have been already created in the laboratory \cite{Terr}.} and represent 
the building blocks for generic graphs. Each point $P$ in the bulk 
$\Gamma \setminus V$ of the graph $\Gamma$ 
belongs to some edge $E_i$ and can be parametrized by the pair of coordinates $(x,i)$, 
where $x>0$ is the distance of $P$ from the vertex (junction) $V$ along $E_i$. 
The embedding of $\Gamma$ and the relative position of 
its edges in the ambient space are irrelevant in what follows. 

This contribution reviews some previous works \cite{Bellazzini:2006jb}, \cite{Bellazzini:2006kh} 
on QFT on star graphs, adding some new results about the spin transport 
and its interplay with the conductance. More details about the conductance away of criticality 
are also provided.  We show that besides the resistance, in this regime the quantum wire 
develops a non-trivial inductance as well. 

The paper is organized as follows. 
In section 2 we describe some specific properties of quantum fields 
on graphs related to symmetries and boundary conditions. In section 3 we 
illustrate the basic concepts developing the theory of scalar fields and vertex 
operators on star graphs. Special attention is devoted to 
the scale invariant boundary conditions, which define in the physical 
terminology the so called critical points of 
the system. The Casimir energy density on 
$\Gamma$ and the correction to the Stefan-Boltzmann law, due to the interaction 
at the vertex $V$, are also determined. In section 4 we describe the bosonization 
procedure on $\Gamma$. We introduce vertex operators and derive the associated 
two-point correlators and scaling dimensions. The results obtained so far are applied in section 5 
for solving the Tomonaga-Luttinger model on $\Gamma$ and answering some basic 
questions left open in the pioneering work by Nayak and collaborators \cite{NFLL}. 
We derive here both the charge and spin conductance and establish a simple relationship 
among them. Section 6 contains our conclusions and some ideas for further developments. 

%\vskip 0.5truecm 

We conclude this short introduction by commenting on the basic tools adopted in the paper. 
Besides the conventional methods of QFT, we use below some elementary results from the spectral theory 
of linear operators on graphs \cite{BCFK}, \cite{E}-\cite{Fulling}, \cite{H1}, \cite{H2}, 
\cite{Kostrykin:1998gz}-\cite{K3}, which is the main topic of the present 
volume. For the application to QFT it is useful to translate this analytic input in 
algebraic form. Since the vertex $V$ of $\Gamma$ can be fully characterized by a scattering matrix, 
it can be considered as a point-like defect (impurity) and one can apply the language 
of reflection-transmission (RT) algebras, developed recently 
\cite{Liguori:1996xr}, \cite{Liguori:1997vd}, \cite{Mintchev:2001aq}-\cite{Mintchev:2007qt} 
in the context of QFT with defects. We will see that RT algebras define a very convenient 
coordinate system in field space in the case of graphs as well.

\section{Characteristic features of QFT on $\Gamma$} 

It is well known that symmetries play a fundamental role in QFT. 
Quantum fields on graphs present in this respect some new features. 
Let $\{j_\nu (t,x,i)\, :\, \nu =t,x\}$ be a conserved current, i.e. 
\be 
\prt_t j_t(t,x,i) - \prt_x j_x(t,x,i)=0 \, . 
\label{curr0}
\ee
In the spirit of axiomatic QFT we assume \cite{KRS} that $j_\nu$ are (tempered) 
operator-valued distributions with a common dense domain $\mathcal D$ in the Hilbert space 
$\mathcal H$ of states of the system.  
The time derivative of the corresponding charge is  
\be 
\prt_t \sum_{i=1}^n \int_0^\infty \rd x\, j_t(t,x,i) = 
\sum_{i=1}^n \int_0^\infty \rd x\, \prt_x j_x(t,x,i) = - \sum_{i=1}^n  j_x(t,0,i)\, ,
\label{charge0}
\ee 
implying charge conservation if and only if Kirchhoff's rule  
\be 
\sum_{i=1}^n  j_x(t,0,i) = 0 
\label{kirgen}
\ee 
holds in the vertex $V$. 

The condition (\ref{kirgen}) 
imposes some restrictions on the interaction at the junction. 
This is a simple but essential novelty with respect to 
QFT on the line $\RR$. Since different conserved currents generate 
in general nonequivalent Kirchhoff's rules, one may expect the presence of obstructions for 
lifting all symmetries on the line to symmetries on $\Gamma$. It may happen in fact that 
two Kirchhoff's rules are in contradiction for {\it generic} boundary conditions. In this case one 
can preserve on $\Gamma$ one of the corresponding symmetries, but not both of 
them\footnote{This feature resembles very much 
anomalies in QFT and will be illustrated explicitly in the next section. 
In our context the phenomenon has its origin in the boundary conditions at the vertex $V$, 
which do not respect in general all symmetries of the system on the line.}. 
One should keep in mind however that for star graphs with even number of edges $n=2k$, 
there exist boundary conditions\footnote{An explicit example is provided in section 3 by the 
$S$-matrix (\ref{scnk}).} for which, after changing the orientation of $k$ edges,  
the graph is equivalent to a bunch of $k$ independent lines. Clearly, for such 
boundary conditions, called in what follows {\it exceptional}, the symmetry content 
of the theory on $\Gamma$ coincides with that on the line. 

The above discussion shows that the choice of boundary conditions in QFT on graphs is a 
crucial point. It is important to have in this respect some general criterion for selection. This is the 
question we are going to address now. We consider in what follows systems which are invariant 
under time translations. It is well known that the relative conserved current is the energy 
momentum tensor $\{\theta_{tt}(t,x,i), \theta_{tx}(t,x,i)\}$. The associated Kirchhoff's rule 
\be 
\sum_{i=1}^n  \theta_{tx}(t,0,i) = 0 
\label{kirteta}
\ee 
selects all boundary conditions for which the Hamiltonian 
\be 
H = \sum_{i=1}^n  \int_0^\infty \rd x\, \theta_{tt}(t,x,i) 
\label{H1}
\ee 
is time independent. Let us observe in passing that for $n=1$ (\ref{kirteta}) implies 
\be 
\theta_{tx}(t,0) = 0 \, , 
\label{kirtetabcft}
\ee 
which is the starting point \cite{Cardy:1984bb} 
of boundary conformal field theory on $\RR_+$. In that case 
one has total reflection from the boundary $x=0$. For $n \geq 2$ partial reflection and transmission 
among the various edges is possible as well. 

Once we have a time-independent Hamiltonian, assuming that the components 
$\{\theta_{tt}(t,x,i), \theta_{tx}(t,x,i)\}$ of the nergy-momentum tensor are symmetric operators on 
$\mathcal D$, one can deduce \cite{KRS} that $H$ is symmetric as well. In order to ensure 
{\it unitary time evolution} (in particular, no dissipation), one must show in addition that $H$ 
admits self-adjoint extensions. Equivalently \cite{RS}, one should check that the deficiency 
indices of $H$ coincide, i.e. 
\be
n_+(H) = n_-(H) \, .  
\label{defind}
\ee
This is a delicate point because (\ref{defind}) is usually hard to verify. Using a well known 
argument \cite{RS} of von Neumann however, the index condition (\ref{defind}) 
is satisfied for systems which are invariant 
under time-reversal. In that case, there exists an {\it antilinear} operator T, which implements 
time-reversal according to 
\be 
T\, H \, T^{-1} = H \, . 
\label{timerev} 
\ee 
In the sequel we consider systems which are invariant under time-reversal. As expected,  
this requirement leads to some restrictions on the boundary conditions. 

In the next section we elaborate on the above points using concrete examples.

\section{Scalar fields on $\Gamma$} 

\subsection{The field $\ph$} 

The massless scalar field $\ph$ on $\Gamma$ is defined by the equation of motion 
\be
\left (\prt_t^2 - \prt_x^2 \right )\ph (t,x,i) = 0\, , 
\qquad x> 0 \, ,  \;  i=1,...,n \, , 
\label{eqm1}
\ee 
the initial conditions (canonical commutation relations) 
\be
[\ph (0,x_1,i_1)\, ,\, \ph (0,x_2,i_2)] = 0\, , \qquad 
\label{initial1}
\ee
\be 
[(\prt_t\ph )(0,x_1,i_1)\, ,\, \ph (0,x_2,i_2)] = 
-\ri \delta_{i_1i_2}\, \delta (x_1-x_2) \, ,
\label{initial2}
\ee 
and the vertex boundary condition 
\be 
\sum_{j=1}^n \left [A_{ij}\, \ph (t,0,j) + B_{ij} (\prt_x\ph ) (t,0,j)\right ] = 0\, , 
\qquad \forall \, t\in \RR\, , \; i=1,...,n\, , 
\label{bc} 
\ee 
$A$ and $B$ being in general two $n\times n$ complex matrices. Clearly the pairs 
$\{A,\, B\}$ and $\{CA,\, CB\}$, where $C$ is any invertible matrix, 
define equivalent boundary conditions. 
The results of \cite{Kostrykin:1998gz} imply that the operator 
$-\prt_x^2$ on $\Gamma$ is self-adjoint provided 
that\footnote{We denote by ${}^*$ the Hermitian conjugation.} 
\be 
A\, B^* - B\, A^* = 0 \,  ,
\label{cond0} 
\ee
and the composite matrix $(A,B)$ has rank $n$. It turns out that these conditions 
ensure also the Kirchhoff's rule for the energy-momentum tensor of $\ph$ 
given by equations (\ref{emtph1}, \ref{emtph2}) below. The general solution of the constraints 
(\ref{cond0}) on $\{A,\, B\}$ is \cite{H1}-\cite{KoS} 
\be
A=C(\II-U)\,, \qquad B=-\ri C(\II+U) \, ,
\label{harm}
\ee
where $C$ is an arbitrary invertible matrix and $U$ is any unitary matrix. 

The problem of quantizing (\ref{eqm1}) with the initial conditions (\ref{initial1}, \ref{initial2}) and the 
boundary condition (\ref{bc}) has a unique solution. It can be written in the form \cite{Bellazzini:2006jb} 
\be
\ph (t,x,i) = \int_{-\infty}^{\infty} \frac{\rd k}{2\pi \sqrt
{2|k|}}
\left[a_i^\ast (k) \e^{\ri (|k|t-kx)} +
a_i (k) \e^{-\ri (|k|t-kx)}\right ] \,  , 
\label{sol1}
\ee
where $\{a_i(k),\, a^*_i(k)\, :\, k\in \RR\}$ generate the 
reflection-transmission (boundary) algebra \cite{Liguori:1996xr}-\cite{Mintchev:2005rz} 
corresponding to the boundary condition (\ref{bc}). This is an associative algebra $\mathcal{A}$ 
with identity element $\bf 1$, whose generators $\{a_i(k),\, a^{* i}(k)\, :\, k\in \RR\}$ 
satisfy the commutation relations  
\bea
&a_{i_1}(k_1)\, a_{i_2}(k_2) -  a_{i_2}(k_2)\, a_{i_1}(k_1) = 0\,  ,
\label{ccr1} \\
&a^\ast_{i_1}(k_1)\, a^\ast_{i_2}(k_2) - a^\ast_{i_2}(k_2)\,
a^\ast_{i_1}(k_1) = 0\,  ,
\label{ccr2} \\
&a_{i_1}(k_1)\, a^\ast_{i_2}(k_2) - a^\ast_{i_2}(k_2)\,
a_{i_1}(k_1) = 
2\pi \left [\delta_{i_1 i_2} \delta(k_1-k_2) +
S_{i_1 i_2}(k_1) \delta(k_1+k_2)\right ] {\bf 1}\,  , 
\nonumber \\ 
\label{ccr3}
\eea 
and the constraints 
\be
a_i(k) = \sum_{j=1}^n S_{ij} (k) a_j (-k) \, , \qquad 
a^\ast_i (k) = \sum_{j=1}^n a^\ast_ j(-k) S_{ji} (-k)\, .    
\label{constr1}
\ee 
The $S$-matrix in (\ref{ccr3}, \ref{constr1}) equals \cite{Kostrykin:1998gz} 
\be 
S(k) = -[(\II - U) + k(\II+U )]^{-1} [(\II - U) - k(\II+U )] 
\label{S1}
\ee
and has the following simple physical interpretation: the diagonal 
element $S_{ii}(k)$ is the reflection amplitude on the edge $E_i$, 
whereas $S_{ij}(k)$ with $i\not=j$ is the transmission amplitude 
from $E_i$ to $E_j$. Being fully characterized by the $S$-matrix (\ref{S1}), 
the vertex $V$ can be viewed as a sort of point-like defect. 
It is not surprising therefore that the algebra $\mathcal{A}$, appearing in the 
context of QFT with boundaries 
or defects \cite{Liguori:1996xr}-\cite{Mintchev:2005rz}, applies in the 
present case as well. $\mathcal{A}$ provides a simple algebraic description of 
the boundary value problem at hand and defines convenient coordinates 
in field space. 

We assume in what follows that $U$ is such that 
\be
\int_{-\infty}^{\infty} \frac{\rd k}{2\pi } \e^{\ri kx} S_{ij} (k) = 0\, , 
\qquad x>0\, . 
\label{compl1}
\ee 
This condition guarantees the absence of bound states and implies the 
canonical commutation relations (\ref{initial1},\ref{initial2}). 

By construction (\ref{S1}) is unitary 
\be 
S(k)^*=S(k)^{-1} \, , 
\label{unit1}
\ee 
and satisfies Hermitian analyticity\footnote{This property implies that the ${}^*$-operation is an 
involution in $\mathcal{A}$.} 
\be 
S(k)^*=S(-k)\, .  
\label{Ha}
\ee 
Combining (\ref{unit1}) and (\ref{Ha}), one gets 
\be 
S(k)\, S(-k) = \II  \, , 
\label{unit2} 
\ee
which ensures the consistency of the constraints (\ref{constr1}). 

The time-reversal transformation (\ref{timerev}) is implemented on $\ph$ by 
\be 
T\ph(t,x)T^{-1} = \ph(-t,x)\, . 
\label{timerev1}
\ee 
Requiring time-reversal invariance and Hermiticity $\ph (t,x,i)^* = \ph (t,x,i)$ one gets \cite{Bellazzini:2006kh} 
\be 
S(k)^t=S(k)\, ,  
\label{symm}
\ee 
where the superscript $t$ denotes transposition. 

Let us focus now on scale invariance. The scaling transformations 
are defined by 
\be 
t \longmapsto \rho t\, , \qquad x \longmapsto \rho x \, , \qquad \rho >0\, .  
\label{sc} 
\ee 
It is well known that systems which are invariant under (\ref{sc}) greatly 
simplify but still capture some universal physical information. 
So, it is interesting to investigate in our context the scale invariant $S$-matrices. 
For this purpose we observe that (\ref{sc}) induce in momentum space  
\be 
k \longmapsto \rho^{-1} k\, .   
\label{sc1} 
\ee 
Accordingly, our system is scale invariant if and only if 
\be 
S(k) = S(\rho^{-1} k) \quad \forall k \in \RR\, , 
\label{sc2}
\ee
which, combined with (\ref{S1}), implies that $S$ is $k$-independent. Therefore, 
in view of (\ref{unit1}, \ref{Ha}, \ref{symm}), any scale invariant $S$-matrix obeys (\ref{compl1}) and 
\be 
S^*=S^{-1}\, ,\qquad S^*=S\, , \qquad S^t=S\, . 
\label{sc3}
\ee 

The classification of these $S$-matrices (critical points) is now a simple 
matter. Indeed, one easily deduces from (\ref{sc3}) that 
the eigenvalues of $S$ are $\pm 1$. Let us denote by $p$ the number of eigenvalues $-1$. 
The values $p=0$ and $p=n$ correspond to the familiar 
Neumann ($S_N=\II$) and Dirichlet ($S_D=-\II$) boundary 
conditions respectively. For $0<p<n$ 
the $S$-matrices satisfying (\ref{sc3}) depend on $p(n-p)\geq 1$ parameters. 

It is instructive to give at this stage some examples. For $n=2$ and $p=1$ one has 
the one-parameter family \cite{Bachas:2001vj}, \cite{Mintchev:2005rz} 
\begin{equation}
S = \frac{1}{1+\alpha^2}
\left(\begin{array}{cc}
\alpha^2-1&-2\alpha\\ 
-2\alpha&1-\alpha^2
\end{array}\right)\, , \qquad \alpha \in \RR\, .  
\label{scnk}
\end{equation}  
The value $\alpha = -1$ illustrates the concept of exceptional boundary conditions, 
introduced in the previous section. One has 
in fact full transmission and no reflection and after changing the orientation of one 
of the edges one gets the free massless scalar field on the line. 

The case $n=3$ has a richer structure: for $p=2$ and 
$p=1$ one has the two-parameter $(\alpha_{1,2} \in \RR)$ families 
\begin{equation} 
S_2(\alpha_1,\alpha_2) =\frac{1}{1+\alpha_1^2 +\alpha_2^2} 
\left(\begin{array}{ccc}
\alpha_1^2-\alpha_2^2 -1&2\alpha_1 \alpha_2 &2\alpha_1 \\ 
2\alpha_1\alpha_2&-\alpha_1^2 +\alpha_2^2 -1&2\alpha_2\\
2\alpha_1&2\alpha_2&1-\alpha_1^2-\alpha_2^2  
\end{array}\right)
\label{famp2}
\end{equation} 
and 
\begin{equation} 
S_1(\alpha_1,\alpha_2) = - S_2(\alpha_1,\alpha_2) \, .   
\label{famp1}
\end{equation} 
Details about the case $n=4$ can be found in \cite{Bellazzini:2006kh}. 

\subsection{The dual field $\phd$ and symmetries} 

The dual field $\phd$ is defined in terms of $\ph$ by the relations 
\be 
\prt_t \phd (t,x,i) = - \prt_x \ph (t,x,i)\, , \quad  
\prt_x \phd (t,x,i) = - \prt_t \ph (t,x,i)\, ,
\qquad x> 0 \, ,  \;  i=1,...,n \, ,  
\label{eqm2}
\ee 
which imply that 
\be
\left (\prt_t^2 - \prt_x^2 \right )\phd (t,x,i)= 0\, , 
\qquad x> 0 \, ,  \;  i=1,...,n  
\label{eqm3}
\ee
as well. The solution is 
\be
\phd (t,x,i) = \int_{-\infty}^{\infty} 
\frac{\rd k\, \varepsilon (k)}{2\pi \sqrt {2|k|}} 
\left[a^\ast_i(k) \e^{\ri (|k|t-kx)} +
a_i (k) \e^{-\ri (|k|t-kx)}\right ] \,  ,  
\label{sol2}
\ee 
where $\varepsilon (k)$ is the sign function. Both $\ph$ and $\phd$ are local fields, 
but we stress that they are not relatively local. This feature is fundamental for bosonization. 

The invariance of the equations of motion (\ref{eqm1},\ref{eqm2}) 
under time translations implies the conservation of the energy-momentum tensor 
\be 
\theta_{tt}(t,x,i) = \frac{1}{2}: \left [ (\prt_t \ph )(\prt_t \ph ) - 
\ph (\prt_x^2 \ph ) \right ]:(t,x,i)\, , 
\label{emtph1}
\ee
\be
\theta_{tx}(t,x,i) = \frac{1}{2} :\left [ (\prt_t \ph)( \prt_x \ph) - 
\ph (\prt_t \prt_x \ph )\right ] :(t,x,i)\, , 
\label{emtph2}
\ee
where $:\cdots :$ denotes the normal product in the algebra $\mathcal A$. 
The associated Kirchhoff rule (\ref{kirteta}) 
is satisfied by construction, being a consequence \cite{Kostrykin:1998gz} of 
(\ref{bc}), (\ref{harm}). 

Equations (\ref{eqm1}, \ref{eqm2}) are also invariant under 
the transformations 
\be 
\ph (t,x,i) \longmapsto \ph (t,x,i) + c\, , \qquad 
\phd (t,x,i) \longmapsto \phd (t,x,i) + {\widetilde c}\, , \quad c,\, {\widetilde c} \in \RR \, ,
\label{shiftsymm}
\ee
which implies the conservation of the currents 
\be 
j_\nu (t,x,i) = \prt_\nu \ph (t,x,i)\, , \qquad 
\jt_\nu (t,x,i)= \prt_\nu \phd (t,x,i)\, , \qquad \nu = t,\, x\, . 
\label{curr1}
\ee 
These currents have a deep physical meaning. In fact, 
in the framework of bosonization $j_\nu$ and $\jt_\nu$ 
control the charge and spin transport respectively. It is therefore 
crucial to check the relative Kirchhoff's rules. 
Using the solution (\ref{sol1}) and the constraints (\ref{constr1}) one finds 
that $j_x$ satisfies Kirchhoff's rule if and only if 
\be 
\sum_{j=1}^n S_{ij}(k) = 1 \, ,\qquad \forall \; i=1,...,n\, , \; k\in \RR \, .  
\label{kir2}
\ee 
Analogously, for $\jt_x$ one finds 
\be 
\sum_{j=1}^n S_{ij}(k) = -1 \, ,\qquad \forall \; i=1,...,n\, , \; k\in \RR \, ,  
\label{kir4}
\ee 
showing that the Kirchhoff's rules for $j_\nu$ and $\jt_\nu$ cannot be satisfied 
simultaneously. Accordingly, one should expect that at most one of the charges 
\be 
Q = \sum_{i=1}^n \int_0^\infty \rd x\, j_t(t,x,i)\, , \qquad 
{\widetilde Q} =  \sum_{i=1}^n \int_0^\infty \rd x\, \jt_t(t,x,i)\, ,
\label{charges}
\ee 
is $t$-independent. This is always the case for $n= 2k+1$ and for generic 
boundary conditions when $n=2k$. Only in the case $n=2k$ with exceptional boundary 
conditions there exist $Q$ and ${\widetilde Q}$ which are both conserved\footnote{Some care is 
needed in the construction of ${\widetilde Q}$ in this case because $\jt_\nu$ is sensitive 
to the orientation of the edges.}. 

The conditions (\ref{kir2}) and (\ref{kir4}) have a simple impact on the scale invariant 
$S$-matrices. Requiring (\ref{kir2}), one eliminates the Dirichlet point ($p=n$) 
and is left only with $p(n-p-1)$ parameters for $0<p<n$. For instance, imposing (\ref{kir2}) 
on (\ref{famp2}), one gets $\alpha_1=\alpha_2=1$, leading to the isolated critical point
\begin{equation} 
S_2 =\frac{1}{3} 
\left(\begin{array}{ccc}
-1&2&2\\ 
2&-1&2\\
2&2&-1  
\end{array}\right) \, ,  
\label{scnay}
\end{equation}
which is invariant under edge permutations.  
{}From (\ref{famp1}) one obtains instead $\alpha_2 = -(1+\alpha_1)$. Therefore, setting  
$\alpha \equiv \alpha_2$, one has in this case the one-parameter family 
\begin{equation} 
S_1(\alpha ) =\frac{1}{1+\alpha +\alpha^2} 
\left(\begin{array}{ccc}
-\alpha&\alpha(\alpha +1)&1+\alpha\\ 
\alpha(\alpha +1)&\alpha +1&-\alpha\\
\alpha +1&-\alpha&\alpha(\alpha +1)  
\end{array}\right)\, , \qquad \alpha \in \RR\, ,   
\label{scnew}
\end{equation} 
which is not invariant under edge permutations for generic $\alpha$. 

Analogous considerations apply to the case when, instead of the Kirchhoff's rule (\ref{kir2}), one 
imposes (\ref{kir4}). In conclusion, we see that the currents $j_\nu$ and $\jt_\nu$ nicely 
illustrate the obstructions which appear for generic boundary conditions when symmetries on 
$\RR$ are lifted to $\Gamma$. 

\subsection{Representations of the RT algebra $\mathcal A$} 

{}For the physical applications one must choose at this point a representation of 
$\mathcal A$. We focus here on two of them: 
the Fock representation $\mathcal F(A)$ and the Gibbs representation $\mathcal G_\beta(A)$ 
at inverse temperature $\beta \sim 1/T$. These two representations have different physical 
interpretation. The fundamental (cyclic) state of $\mathcal F(A)$ does not 
contain particle excitations (vacuum state) and is annihilated by the generators $a_i(k)$ 
of $\mathcal A$. On the contrary, the fundamental state of $\mathcal G_\beta(A)$ 
involves an infinite number of particles, which model a heat bath maintaining the system 
at constant inverse temperature $\beta$. Referring for the details 
concerning $\mathcal F(A)$ and $\mathcal G_\beta(A)$ to \cite{Liguori:1996xr} and 
\cite{Mintchev:2004jy}, we collect below only the two-point expectation values of 
$\{a_i(k),\, a^*_i(k)\, :\, k\in \RR\}$. 
In $\mathcal F(A)$ one has 
\be
\langle a_i(p)a^\ast_j(q)\rangle  = 
2\pi \left [\delta_{ij}\, \delta (p-q) 
+ S_{ij}(p)\, \delta (p+q)  \right ]\,  , \qquad 
\langle a^\ast_i (p )a_j(q)\rangle = 0\, .   
\label{fock1}
\ee
In $\mathcal G_\beta(A)$ one finds instead 
\bea 
\langle a_i(p)a^\ast_j(q)\rangle_\beta  =
\frac{1}{ 1 - \e^{-\beta |p|}} \, 
2\pi \left[\delta_{ij} \delta (p-q) 
+ S_{ij}(p)\delta (p+q)  \right ]\,  ,
\label{gibbs1}\\
\qquad \; \langle a^\ast_i(p)a_j(q)\rangle_\beta  =
\frac{\e^{-\beta |p|}}{ 1 - \e^{-\beta |p|}}\,  
2\pi \left [ \delta_{ij} \delta (p-q) 
+ S_{ij}(-p)\delta (p+q)  \right ] \, . 
\label{gibbs2}
\eea 

As an application of the Gibbs representation $\mathcal G_\beta(A)$, we derive now 
the energy density $\langle \theta_{tt} (t,x,i) \rangle_{\beta}$ in the Gibbs state. 
This quantity is determined up to a constant. Taking as a reference point the energy density 
on the whole line $\RR$ and at zero temperature and using the point-splitting regularization, 
one finds \cite{Bellazzini:2006jb} 
\be 
\langle \theta_{tt} (t,x,i) \rangle_{\beta} - 
\langle \theta_{tt} (t,x) \rangle_{\infty}^{\rm line} = 
\varepsilon_{{}_{\rm S-B}} (\beta) + \E_{{}_{\rm C}} (x,i) + \E(x,i,\beta ) \, , 
\label{endens3}
\ee 
where 
\be 
\varepsilon_{{}_{\rm S-B}} (\beta) = 
\int_{-\infty}^{+\infty}\frac{\rd k}{2\pi}\, |k| 
\;\frac{e^{-\beta |k|}}{1 - e^{-\beta |k|}}  =  \frac{\pi}{6 \beta^2} \sim T^2\, , 
\label{sb4}
\ee 
is the Stefan-Boltzmann contribution, 
\be
\E_{{}_{\rm C}} (x,i) = \frac{1}{2} \int_{-\infty}^{+\infty}\frac{\rd k}{2\pi}\, |k|\,
S_{ii}(k)e^{2\ri kx}   
\label{vc2}
\ee 
is the Casimir energy density associated with the interaction in the vertex of $\Gamma$ 
at zero temperature and 
\be
\E (x,i,\beta) = \int_{-\infty}^{+\infty}\frac{\rd k}{2\pi}\, |k| 
\;\frac{e^{-\beta |k|}}{1- e^{-\beta |k|}}\, S_{ii}(k)e^{2\ri kx}\,  
\label{vc3}
\ee 
is the finite temperature correction to the latter. The integrals over $k$ are easily 
computed in the scale invariant case and give  
\be
\E_{{}_{\rm C}} (x,i) = -\frac{S_{ii}}{8\pi x^2} \, , \qquad 
\E (x,i,\beta) =  \frac {\pi S_{ii}}{2 \beta^2\, {\sinh}^2 \left (2\pi\frac{x}{\beta} \right )} 
-\frac{S_{ii}}{8\pi x^2} \, .  
\label{vc5}
\ee 
The Casimir energy of a segment $[\delta \, ,\, \delta +L]\subset E_i$ with $\delta,\, L>0$ 
is therefore 
\be 
E_{{}_{\rm C}} (i;\delta, L) = \int_\delta^{\delta+L} \rd x\, \E_{{}_{\rm C}} (x,i) 
= -\frac{S_{ii}}{8\pi} \frac{L}{\delta (\delta +L)} \, , 
\label{vc6}
\ee 
which is increasing with the length $L$ (for any $\delta>0$) if $S_{ii} <0$ and decreasing if 
$S_{ii} >0$. For the properties of the Casimir energy on star graphs 
with compact edges we refer to \cite{Fulling}.

\section{Bosonization and vertex operators} 

Bosonization is a fundamental tool of quantum field theory on the line. 
It plays an essential role for constructing exact solutions \cite{Lowenstein:1971fc}, 
\cite{Hald}, \cite{Voit} of some models and establishing the equivalence among others \cite{Coleman:1974bu}. 
The non-abelian variant of bosonization \cite{Witten:1983ar} produces also relevant results 
in two-dimensional conformal field theory and string theory. Bosonization is in particular 
the basis of the Coulomb gas representation \cite{DiFrancesco:1997nk} of the minimal 
conformal invariant models on the line. 

The main idea of bosonization belongs to Jordan and Wigner and 
dates back to 1928. There exist nowadays a rather extensive literature on the subject 
(see e.g. \cite{S} and references therein). It is well-known 
that the fundamental building blocks on the line are the free massless scalar field 
$\ph $ and its dual $\phd $. Both $\ph$ and $\phd $ are {\it local} fields, 
but it turns out that they are not {\it relatively local}. 
As recognized already in the early sixties \cite{Wightman}, this feature is in the heart 
of bosonization, duality, chiral superselection sectors, generalized (anyonic) statistics  
and many other phenomena, characterizing the rich 
structure of QFT on the line. We already mentioned that the fields $\ph$ and $\phd$, 
defined by (\ref{sol1}) and (\ref{sol2}), preserve the above mentioned locality properties on 
$\Gamma$ as well. It is quite natural therefore to extend \cite{Bellazzini:2006kh} the 
bosonization procedure on the line to star graphs. The extension involves 
composite operators which are constructed in terms of $\ph$ and $\phd $ and have non-trivial 
scaling dimensions and peculiar exchange relations. In particular, interacting fermions 
can be expressed in this framework via the free bosonic fields $\ph$ and $\phd $ 
(hence the name ``bosonization''). Due to the point-like interaction at the vertex $V$, 
the construction on $\Gamma$ presents new intriguing features, which are briefly 
summarized below. 

\subsection{Vertex operators}

Following \cite{Wightman}, the basic objects constructed in terms of $\ph$ and $\phd$ 
are the so called vertex operators, which are essentially exponentials of $\ph$ and $\phd$. 
The vertex operators\footnote{See for example \cite{Kac} for more details about these operators on the line.} 
allow one to represent fermion fields and, more generally anyons \cite{Leinaas:1977fm}, in terms 
of boson fields. Using the vertex representation of fermion fields, one can express the 
conserved vector and axial fermion currents as gradients of $\ph$ and $\phd$ respectively, 
which greatly simplifies the study of the current-current interactions. 
We will illustrate all these aspects here and in the next section. 

Although vertex operators exist \cite{Mintchev:2005rz} on $\Gamma$ for the general 
boundary condition (\ref{bc}), at a scale invariant point they have 
simpler and more remarkable structure. For this reason we focus to the end of the paper 
on the scale invariant case and introduce for convenience the right and left chiral fields
\be 
\ph_{i, R} (t-x)=\ph(t,x,i)+\phd(t,x,i)\, , \qquad 
\ph_{i,L}(t+x)=\ph(t,x,i)-\phd(t,x,i)\, . 
\label{rlbasis}
\ee 
Inserting (\ref{sol1},\ref{sol2}) in (\ref{rlbasis}) one gets 
\be
\ph_{i,R}(\xi ) = \int_{0}^{\infty} \frac{\rd k}{\pi \sqrt
{2k}}
\left[a^\ast_i (k) \e^{\ri k\xi} +
a_i (k) \e^{-\ri k\xi}\right ] \,  , 
\label{pir}
\ee
which satisfy 
\begin{equation}
[\ph_{i,R} (\xi _1 )\, ,\, \ph_{j,R}(\xi _2 )] = -\ri \varepsilon (\xi _{12})\delta_{ij} \, , 
\qquad  \xi _{12} \equiv \xi_1-\xi_2\, .
\label{commr}
\end{equation}
Moreover, using that $S$ is constant by scale-invariance, the constraints (\ref{constr1}) imply 
\begin{equation}
\ph_{i,L} (\xi) = \sum_{j=1}^n S_{ij}\, \ph_{j,R}(\xi ) \, . 
\label{pil}
\end{equation} 
We will need also the chiral charges given by
\be 
Q_{i,Z} = \frac{1}{4} \int_{-\infty}^{\infty} \rd \xi \, \prt_\xi \ph_{i,Z} (\xi)\, ,  
\qquad Z=R,\, L\, . 
\label{lrcharges}
\ee 
Using (\ref{ccr1}-\ref{ccr3}), one can easily verify that 
\be 
[Q_{i_1,Z_1}\, ,\, Q_{i_2,Z_2}]=0\, ,  
\label{qcomm1}
\ee  
\be 
[Q_{i_1,Z_1}\, ,\, \ph_{i_2,Z_2}(\xi)] = \left\{\begin{array}{cc}
-\frac{\ri}{2} \delta_{i_1i_2} \, ,
& \quad \mbox{$Z_1=Z_2$}\, ,\\[1ex]
\, -\frac{\ri}{2} S_{i_1i_2}\, ,
& \quad \mbox{$Z_1\not=Z_2$}\, . \\[1ex]
\end{array} \right.
\label{qcomm2}
\ee 

Now, we are in position to introduce a family of vertex operators 
parametrized by $\zeta = (\sigma , \tau) \in \RR^2$ and defined by 
\be  
v(t,x,i;\zeta) = z_i\, q(i;\zeta) 
:\exp\left \{\ri \sqrt{\pi}\left [\sigma \ph_{i,R}(t-x) + 
\tau \ph_{i,L}(t+x)\right]\right \}: \, , 
\label{vertex1}
\ee 
where the value of the normalization constant $z_i\in \RR$ will be fixed later on,  
\be 
q(i;\zeta)= \exp\left [\ri \sqrt{\pi}\left (\sigma Q_{i,R} -\tau Q_{i,L}\right )\right ]\, ,  
\label{qfact}
\ee 
and $: \cdots :$ denotes the normal product in the algebra $\mathcal{A}$. 

The exchange properties of $v(t,x,i;\zeta )$ 
determine their statistics. A standard calculation shows that 
\bea
v(t_1,x_1,i_1;\zeta_1 ) v(t_2,x_2,i_2;\zeta_2 ) = \qquad \qquad \qquad 
\nonumber \\
{\mathcal R} (t_{12},x_1,i_1,x_2,i_2; \zeta_1,\zeta_2 )\, 
v(t_2,x_2,i_2;\zeta_2 ) v(t_1,x_1,i_1;\zeta_1 ) \, , 
\label{exch}
\eea
where the exchange factor $\mathcal R$ is a c-number and $t_{12} \equiv t_1-t_2$. 
The statistics of $v(t,x,i;\zeta )$ is determined by the 
value of $\mathcal R$ at space-like separation $t_{12}^2-x_{12}^2 <0$ with $x_{12} \equiv x_1-x_2$.   
By means of equations (\ref{commr}, \ref{pil}, \ref{qcomm1}, \ref{qcomm2}) one finds  
\be 
{\mathcal R}(t_{12}, x_1,i_1, x_2,i_2;\zeta_1,\zeta_2 ){\Big \vert_{t_{12}^2-x_{12}^2 <0}} = 
\e^{-\ri \pi (\sigma_1 \sigma_2 -\tau_1 \tau_2) \varepsilon (x_{12}) \delta_{i_1i_2}}
\, . 
\label{exchf1}
\ee 
Therefore $v(t,x,i;\zeta )$ obey anyon (abelian braid) statistics \cite{Liguori:1993pp} with 
statistical parameter 
\be 
\vartheta =  \sigma_1 \sigma_2 -\tau_1 \tau_2 \, ,
\label{statpar}
\ee 
when localized at the same wedge $E_i$. Otherwise, $v(t,x,i;\zeta )$ commute. 

In order to obtain fermions, we take any $\zeta = (\sigma>0, \tau)$ 
with $\sigma \not= \pm \tau$ and set 
\be 
\zeta^\prime = (\tau, \sigma)\, . 
\label{zprime}
\ee 
Then we define 
\be 
\mathcal{V} (t,x,i;\zeta ) = \eta_i\, v(t,x,i;\zeta )\, , 
\qquad \mathcal{V}(t,x,i;\zeta^\prime ) = \eta_i^\prime\, v(t,x,i;\zeta^\prime )\, ,
\label{vertex2}
\ee 
where $\{\eta_i, \eta_i^\prime\}$ are the so called Klein factors \cite{Bellazzini:2006kh} 
satisfying the relations 
\be 
\eta_{i_1} \eta_{i_2} + \eta_{i_2} \eta_{i_1} = 2 \delta_{i_1i_2} {\bf 1}\ , \quad  
\eta_{i_1}^\prime \eta_{i_2}^\prime + \eta_{i_2}^\prime \eta_{i_1}^\prime = 2 \delta_{i_1i_2} {\bf 1}\, ,  \quad 
\eta_{i_1} \eta_{i_2}^\prime + \eta_{i_2}^\prime \eta_{i_1} = 0 \, . 
\label{klein}
\ee 
It is not difficult to show \cite{Bellazzini:2006kh} that (\ref{vertex2}) and their Hermitian conjugates 
obey Fermi statistics provided that 
\be 
 \sigma^2 -\tau^2 = 2k+1\, , \qquad   k\in \ZZ \, . 
 \label{fermi}
 \ee 
 {}For $k=0$ on has in particular canonical fermions. 

\subsection{Correlation functions} 

The basic two point correlators of the chiral fields in the Fock 
representation $\mathcal F(A)$ are easily derived: 
\be 
\langle \ph_{i_1,R}(\xi_1) \ph_{i_2,R}(\xi_2)\rangle = 
\langle \ph_{i_1,L}(\xi_1) \ph_{i_2,L}(\xi_2)\rangle =
\delta_{i_1i_2}\, u(\mu \xi_{12}) \, , 
\label{cf2}
\ee
\be 
\langle \ph_{i_1,R}(\xi_1) \ph_{i_2,L}(\xi_2)\rangle = 
\langle \ph_{i_1,L}(\xi_1) \ph_{i_2,R}(\xi_2)\rangle = S_{i_1i_2}\, u(\mu \xi_{12}) \, ,  
\label{cf6}
\ee 
where 
\be 
u(\mu \xi)=-\frac{1}{\pi} \ln (\mu |\xi|) -\frac{i}{2}\varepsilon (\xi) = 
-\frac{1}{\pi} \ln (i\mu \xi + \epsilon )\, , \qquad \epsilon > 0\, . 
\label{log}
\ee 

In order to compute the correlation functions of the vertex operators (\ref{vertex2}), 
we need also a representation of the algebra of Klein factors. We adopt the one defined by 
the two-point correlators 
\be 
\langle \eta_{i_1} \eta_{i_2} \rangle = 
\langle \eta_{i_1}^\prime \eta_{i_2}^\prime \rangle = 
\langle \eta_{i_1} \eta_{i_2}^\prime \rangle = 
-\langle \eta_{i_2}^\prime \eta_{i_1} \rangle = 
\kappa_{i_1i_2} =
\left\{\begin{array}{cc}
\; \; 1 \, ,
& \quad \mbox{$i_1\leq i_2$}\, ,\\[1ex]
-1\, ,
& \quad \mbox{$i_1>i_2$}\, , \\[1ex]
\end{array} \right.
\label{klein2}
\ee 

A standard computation \cite{Bellazzini:2006kh} now gives  
\bea  
\langle \mathcal{V} (t_1,x_1,i_1;\zeta) \mathcal{V}^*(t_2,x_2,i_2;\zeta) \rangle = 
\qquad \qquad \qquad \qquad \nonumber \\
z_{i_1}z_{i_2}
 \mu^{-[(\sigma^2+\tau^2)\delta_{i_1i_2} +2\sigma \tau S_{i_1i_2}]} 
 \kappa_{i_1i_2}
\left [\frac{1}{\ri (t_{12}-x_{12})+ \epsilon}\right ]^{\sigma^2 \delta_{i_1i_2}} 
\left [\frac{1}{\ri (t_{12}+x_{12})+ \epsilon}\right ]^{\tau^2 \delta_{i_1i_2}}
\quad \nonumber \\
\left [\frac{1}{\ri (t_{12}-{\widetilde x}_{12})+ \epsilon}\right ]^{\sigma \tau S_{i_1i_2}}
\left [\frac{1}{\ri (t_{12}+{\widetilde x}_{12})+ \epsilon}\right ]^{\sigma \tau S_{i_1i_2}}\; \; \; \, 
\label{cf7}
\eea
with ${\widetilde x}_{12}=x_1+x_2$. This equation suggest to take the normalization factor 
\be 
z_i = \mu^{\frac{1}{2}(\sigma^2+\tau^2 + 2\sigma \tau S_{ii})} \, . 
\label{z}
\ee 
In this way the vertex correlator (\ref{cf7}) is $\mu$-independent, when 
localized on the same edge. 
Performing in (\ref{cf7}) the scaling transformation (\ref{sc}), one obtains 
\be 
\langle \mathcal{V} (\rho t_1,\rho x_1,i_1;\zeta) \mathcal{V}^*(\rho t_2,\rho x_2,i_2;\zeta) \rangle = 
\rho^{-D_{i_1i_2}}\, \langle \mathcal{V}(t_1,x_1,i_1;\zeta) \mathcal{V}^*(t_2,x_2,i_2;\zeta) \rangle \, , 
\label{sctransf}
\ee
where 
\be 
D = (\sigma^2 +\tau^2)\II_n + 2\sigma \tau S\, . 
\label{msc}
\ee 
The scaling dimensions $d_i$ are determined by the eigenvalues of the matrix $D$. 
Diagonalizing (\ref{msc}) one finds 
\be 
d_i = \frac{1}{2}(\sigma^2 + \tau^2) + \sigma \tau s_i \, , \qquad i=1,...,n \, ,  
\label{dimensions0}
\ee 
where $s_i$ are the eigenvalues of $S$. Since in the scale 
invariant case $s_i=\pm 1$, one has 
\be 
d_i = \frac{1}{2}(\sigma + s_i\tau)^2 \geq 0 \, .   
\label{dimensions}
\ee 
Recalling that the same vertex operator on the line $\RR$ has dimension 
\be 
d_{\rm line} =  \frac{1}{2}(\sigma^2 + \tau^2) \, ,   
\label{dimline}
\ee 
we see that the interaction at the junction affects the scaling dimensions. 

\section{The four-fermion interaction on $\Gamma$} 

In this section we introduce and investigate non-trivial bulk interactions on $\Gamma$. 
We focus on the Tomonaga-Luttinger (TL) model which captures the universal features of a 
wide class of one-dimensional quantum many-body systems called Luttinger liquids. 
The Lagrangian density defining the dynamics of the TL model is 
\begin{equation} 
\mathcal{L} = \ri \psi_1^*(\prt_t + v_F\prt_x)\psi_1 +  \ri \psi_2^*(\prt_t - v_F\prt_x)\psi_2 
-g_+(\psi_1^* \psi_1+\psi_2^* \psi_2)^2 - g_-(\psi_1^* \psi_1-\psi_2^* \psi_2)^2\, ,  
\label{lagrangian}
\end{equation} 
where $\psi_\alpha = \psi_\alpha (t,x,i)$ with $\alpha =1,2$ are complex fermion fields, 
$v_F$ is the Fermi velocity and $g_\pm \in \RR$ are the coupling constants. 

It is well known \cite{Hald,Voit} that this model is exactly solvable on the line  $\RR$ by bosonization. 
On the graph $\Gamma$ however the situation is more involved, because some boundary conditions 
must be imposed in the vertex $V$. The simplest boundary conditions one can imagine 
are linear in $\psi_\alpha$ and are generated by the variation of 
the boundary Lagrangian density\footnote{$\mathcal{B}$ is 
the matrix defining the boundary interaction among the fermions.} 
\be 
\mathcal{L}_V = \psi^*_\alpha (t,0,i)\mathcal{B}_{\alpha \beta\; ij}\, \psi_\beta (t,0,j) \, , 
\label{blagrangian}
\ee 
localized at $x=0$. After bosonization $\mathcal{L}_V$ involves exponential boundary interactions of the scalar field 
and the theory defined by $\mathcal{L} + \mathcal{L}_V$ is no longer exactly solvable. Nevertheless, using 
instanton gas expansion and strong-weak coupling duality on a star graph with $n=3$ edges, 
Nayak and collaborators \cite{NFLL} established the existence of a critical point in which the electric  
conductance $G$ is enhanced\footnote{A possible explanation \cite{NFLL} of this result 
is based on the so-called Andreev reflection \cite{A}.} with respect to that on the line, namely 
\be 
G = \frac{4}{3}\, G_{\rm line} \, . 
\label{enh1}
\ee
This inspiring result, which has been confirmed more recently by different authors 
\cite{Sch, DRS, Oshikawa:2005fh}, raises some interesting open problems:  

(i) existence of other critical points and their classification; 

(ii) behavior under edge permutations; 

(iii) is enhancement of the conductance universal or reduction 
is possible as well? 

(iv) what is the law governing the spin transport? 

These questions are answered in the rest of the paper. The main idea 
behind our analysis is to modify the boundary conditions in such a way that 
they become linear after bosonization. We show that this is indeed possible\footnote{As expected, 
the new boundary conditions are non-linear in terms of $\psi_\alpha$.} and, applying the results of 
section 3, the model can be solved exactly. All critical points can be classified and their 
characteristic features are easily investigated. 

\subsection{The Thirring model on $\Gamma$ and its solution} 

{}It will be enough for our purposes to analyze the TL model in the special case 
when $g_+=-g_-\equiv g\pi >0$ and $v_F=1$. The classical equations of motion 
of this system, known as a Thirring model, can be written in 
the following matrix form 
\be 
\ri (\gamma_t \prt_t - \gamma_x \prt_x)\psi (t,x,i) = 
2\pi g [\gamma_t J_t(t,x,i) - \gamma_x J_x(t,x,i)]\psi (t,x,i) \, , 
\label{eqmt} 
\ee 
where
\be 
\psi (t,x,i)=\pmatrix{ \psi_1(t,x,i) \cr \psi_2(t,x,i) \cr}\, ,  
\qquad 
\gamma_t = \pmatrix{ 0 & 1 \cr 1 & 0 \cr}\, , \qquad
\gamma_x = \pmatrix{ 0 & 1 \cr -1 & 0 \cr} \, .
\label{gamma}
\ee 
and 
\be 
J_\nu (t,x,i) = \overline \psi (t,x,i) \gamma_\nu \psi (t,x,i) \, , \qquad 
\overline \psi \equiv \psi^\ast \gamma_t \, . 
\label{currt1} 
\ee
is the conserved vector current. The number of edges $n$ is arbitrary. 

{}For quantizing the model, we set $\sigma >0$ and 
\be
\psi_1 (t,x,i) = \frac{1}{\sqrt {2\pi}}\mathcal{V} (t,x,i;\zeta )\, , \qquad 
\psi_2 (t,x,i) = \frac{1}{\sqrt {2\pi}}\mathcal{V} (t,x;\zeta^\prime ) \, .
\label{psit}
\ee 
In order to have canonical fermions we require 
\be 
\sigma^2-\tau^2=1\, ,  
\label{condt}
\ee
implying $\sigma \not= \pm \tau$. 

The quantum current $J_\nu $ is constructed by point-splitting according to 
\be
J_\nu (t,x,i) = \frac{1}{2} \lim_{\epsilon \to +0} Z (\epsilon) 
\left [\, \overline \psi (t,x,i)\gamma_\nu \psi (t, x+\epsilon, i) + 
\overline \psi (t, x+\epsilon, i) \gamma_\nu \psi (t, x, i)\, \right ]\, , 
\label{pointsplit3}  
\ee 
where $Z (\epsilon)$ is some renormalization constant. The latter can 
be fixed \cite{Bellazzini:2006kh} in such a way that 
\be 
J_\nu (t,x,i) = - \frac{1}{(\sigma + \tau)\sqrt \pi}\,  \prt_\nu \ph (t,x,i) 
= - \frac{1}{(\sigma + \tau)\sqrt \pi}\,  j_\nu (t,x,i) \, , 
\label{currt2} 
\ee
thus generating the $U(1)$-charge transformation 
\be 
[J_t (t,x,i)\, ,\, \psi(t,y,j)] = -\delta(x-y) \delta_{ij} \psi (t,x,j) \, . 
\label{wi1}
\ee 
Because of (\ref{currt2}) the quantum equation of motion takes the form 
\be
i(\gamma_t \prt_t - \gamma_x \prt_x)\psi (t,x,i) = 
-\frac{2g\sqrt \pi }{(\sigma + \tau)} : \left (\gamma_t \prt_t \ph - 
\gamma_x \prt_x \ph \right ) \psi : (t,x,i)  \, .  
\label{qeqmt}
\ee
Now, using the vertex realization (\ref{psit}) of $\psi$, one easily 
verifies that (\ref{qeqmt}) is satisfied provided that 
\be 
\tau(\sigma+\tau) = g\, . 
\label{eqmts}
\ee
Combining (\ref{condt}) and (\ref{eqmts}) one determines $\sigma$ and $\tau$ in terms of the coupling 
constant: 
\be 
\sigma = \frac{1+g}{\sqrt {1+2g}} >0\, , \qquad \tau =  \frac{g}{\sqrt {1+2g}} \, . 
\label{solt}
\ee 

We generated above a solution of the Thirring model on $\Gamma$ and it is natural to ask 
what kind of boundary conditions does it satisfy. From (\ref{pil}) and (\ref{currt2}) one immediately finds 
\begin{equation} 
J_x (t,0,i) = -\sum_{k=1}^n S_{ik}\,  J_x (t,0,k)\, , 
\label{currbc}
\end{equation} 
which has a transparent physical interpretation. The boundary condition (\ref{currbc}) 
describes the dissipationless splitting of the charge current at the junction $x=0$. 
According to (\ref{currt1}, \ref{currt2}) it is {\it quadratic} in the fermion field $\psi_\alpha$, but 
{\it linear} in $\ph$. This fundamental difference with respect to the boundary conditions 
adopted in \cite{NFLL} is the crucial novelty allowing to solve the model exactly. 

The scaling dimensions of our solution follow from (\ref{dimensions}). The result is 
\be 
d_i = \frac{1}{2} (1+2g)^{s_i} \, , \qquad s_i = \pm 1 
\ee
and reflects both the non-trivial bulk and boundary (vertex) interactions. 

Let us discuss now the spin-transport. It is well known that there is no true spin in one space dimension (and 
therefore on $\Gamma$) because there are no space-rotations. Nevertheless, one can associate 
a ``spin" $-\frac{1}{2}$ to $\psi_1$ and $\frac{1}{2}$ to $\psi_2$. This assignment 
is not only formal, because there exists a conserved current describing the transport of this 
quantum number (chirality). Consider in fact the axial current 
\be
\Jt_\nu (t,x,i) = \frac{1}{2} \lim_{\epsilon \to +0} {\widetilde Z} (\epsilon) 
\left [\, \overline \psi (t,x,i)\gamma_\nu \gamma_5 \psi (t, x+\epsilon, i) + 
\overline \psi (t, x+\epsilon, i) \gamma_\nu \gamma_5 \psi (t, x, i)\, \right ] 
\label{pointsplit4}  
\ee 
with $\gamma_5=-\gamma_t \gamma_x$. A suitable choice \cite{Bellazzini:2006kh} of the renormalization constant 
${\widetilde Z} (\epsilon)$ leads to 
\be 
\Jt_\nu (t,x,i) = -\frac{1}{2(\sigma - \tau)\sqrt \pi}\,  \prt_\nu \phd (t,x,i) 
= -\frac{1}{2(\sigma - \tau)\sqrt \pi}\, \jt_\nu (t,x,i) \, , 
\label{currt3} 
\ee
which is indeed conserved. Moreover, 
\be 
[\Jt_t (t,x,i)\, ,\, \psi_\alpha (t,y,j)] = 
\left\{\begin{array}{cc}
-\frac{1}{2} \delta(x-y) \delta_{ij} \psi_1 (t,y,j) \, ,
& \quad \mbox{$\alpha =1$}\, ,\\[1ex]
\; \; \; \frac{1}{2} \delta(x-y) \delta_{ij} \psi_2 (t,y,j)\, ,
& \quad \; \mbox{$\alpha = 2$}\, . \\[1ex]
\end{array} \right. 
\label{wi2}
\ee 

We turn finally to the critical points. As already discussed in section 3.1, the critical points are fully classified 
by eq. (\ref{sc3}). If one requires in addition the conservation of the charge associated with the electric 
current (\ref{currt2}), one must impose the Kirchhoff's rule (\ref{kir2}). In this way, besides the Neumann point, 
the remaining critical points for $n=3$ are 
$S_2$ and $S_1(\alpha )$, defined by (\ref{scnay}, \ref{scnew}). We will see in the next section that 
$S_2$ is precisely the critical point discovered by Nayak and collaborators in \cite{NFLL}. The one-parameter family 
$S_1(\alpha )$ is new and is not invariant under edge permutations for generic $\alpha$. This statement 
clarifies points (i) and (ii) at the beginning of this section. 

\subsection{Charge and spin transport} 

In order to derive the electric and spin conductance, we couple the system to a {\it classical} 
external field $A_\nu (t,x,i)$ by means of the substitution 
\be 
\prt_\nu \longmapsto \prt_\nu + \ri A_\nu (t,x,i)  
\label{covder}
\ee 
in eq. (\ref{eqmt}). The resulting Hamiltonian is time dependent. The conductance can be extracted 
from the expectation value $\langle J_x(t,x,i)\rangle_{A_\nu}$, more precisely, from 
the linear term of the expansion of $\langle J_x(t,x,i)\rangle_{A_\nu}$ in terms of $A_\nu$. 
This term can be computed by linear response theory which gives for the 
conductance tensor \cite{Bellazzini:2006kh} 
\be
G_{ij} = G_{\rm line} \left (\delta_{ij}-S_{ij} \right ) \, , 
\label{conduc1}
\ee  
where 
\be 
G_{\rm line} = \frac{1}{2\pi (\sigma +\tau)^2} = \frac{1}{2\pi (1+2g)} 
\label{gline}
\ee
is the conductance on the line $\RR$. 

The simple expression (\ref{conduc1}), describing 
the electric conductance of the Thirring model at a critical point, has a number of 
remarkable properties. As expected, it satisfies the Kirchhoff's rule 
\be
\sum_{j=1}^n G_{ij} =0 \, , \qquad i=1,...,n\, , 
\label{kircond}
\ee
provided that (\ref{kir2}) holds, i.e. the electric charge is conserved. The conductance 
$G_{ii}$ of the edge $E_i$ is enhanced with respect to $G_{\rm line}$ if $S_{ii}<0$ 
and reduced if $S_{ii}>0$. In particular, at the critical point (\ref{scnay}) one reproduces 
the result (\ref{enh1}) of \cite{NFLL}. Note that enhanced conductance takes place when 
the Casimir energy (\ref{vc6}) is increasing with $L$ and vice versa. 

The properties (\ref{sc3}) of the $S$-matrix imply $|S_{ii} | \leq 1$, 
leading to the simple bound 
\be 
0\leq G_{ii} \leq 2 G_{{\rm line}} \, ,  
\label{bound}
\ee 
where we have used that $G_{\rm line}$ is positive. 
Another constraint on the diagonal elements of $G$ is the sum rule 
\be 
{\rm Tr}\, G = 2p\, G_{{\rm line}}\, , 
\label{tr2}
\ee
$p$ being the number of eigenvalues $-1$ of $S$. 

\begin{figure}[h]
\begin{center}
\includegraphics[scale=1]{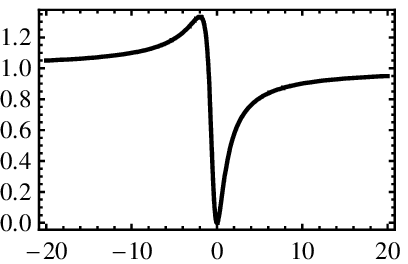}
%\hskip 0.1 truecm 
%\includegraphics[scale=1]{fig-B2.pdf}
\includegraphics[scale=1]{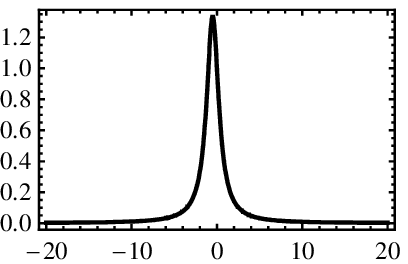}
%\hskip 0.1 truecm 
%\includegraphics[scale=1]{fig-C2.pdf}
\includegraphics[scale=1]{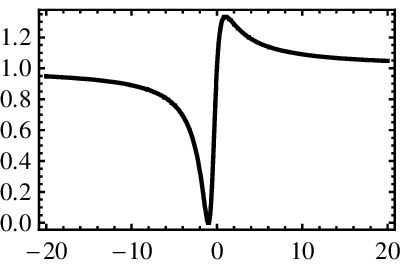}
\end{center}
\caption{$G_{11} (\alpha)$ (left), $G_{22} (\alpha)$ and 
$G_{33} (\alpha)$ (right) for $G_{{\rm line}}=1$.}\label{condfig} 
\end{figure}

Inserting the new family of critical points (\ref{scnew}) in (\ref{conduc1}) we conclude that 
both enhancement and reduction of the conductance with respect to the line are possible. 
This fact is illustrated by the plots in Fig.~\ref{condfig} and answers the question (iii) from the beginning of this section. 

Away of criticality the conductance takes the form \cite{Bellazzini:2006kh}
\be
G_{ij}(\omega) = G_{{\rm line}} \left [\delta_{ij}-S_{ij}(\omega)\right ] \, ,
\label{condomega}
\ee 
where $S$ is given by (\ref{S1}) and $\omega$ is the frequency of the Fourier 
transform ${\widehat A}_x(\omega,i)$ of the external field $A_x(t,i)$ applied to the system. 
Since $S_{ij}(\omega)$ are in general complex, in this regime the system 
admits non-trivial inductance as well. Let us compare for instance a star graph with $n=2$ edges 
to an impedance $Z\left(\omega\right)$, characterized by the condition 
$Z\left(1\right)=R+ \ri L$. 
The ingoing electric currents $I\left(\omega,i\right)=\langle J_x(\omega,i)\rangle_{A_\nu}$ 
can be expressed in the form 
\be
\left( \begin{array}{l}
 I\left(\omega,1\right)  \\ 
 I\left(\omega,2\right)  \\ 
 \end{array} \right) = Z^{ - 1}\left(\omega\right) \left( {\begin{array}{*{20}c}
  \; \; 1 & -1  \\
   -1 &\; \; \, 1 \\ 
   \end{array}} \right)\left( \begin{array}{l}
 {\widehat A}_x(\omega,1)  \\ 
 {\widehat A}_x(\omega,2)  \\ 
 \end{array} \right) \, . 
\label{impedence2}
\ee 
Confronting this relation with (\ref{condomega}) and using the explicit form (\ref{S1}) 
of $S\left(\omega\right)$ one finds 
\be
Z\left(\omega\right)=R+\ri \omega L \, ,
\label{inductance2}
\ee
so that the most general impedance, which can be obtained for $n=2$, 
has a real component $R=G_{{\rm line}}^{\,-1}$ and a generic inductive 
component $\ri \omega L$ depending on the boundary conditions. Following 
the above procedure, one can analyze all the admittances characterizing 
star graphs with $n>2$ away of criticality. 

{}Finally, to answer the last question (iv), we investigate the spin 
transport governed by the dual current (\ref{currt3}). In this case 
one should evaluate the expectation value $\langle \Jt_x(t,x,i)\rangle_{A_\nu}$. 
The result for the spin conductance tensor is 
\be
\Gt_{ij} = \Gt_{\rm line} \left (\delta_{ij}+S_{ij} \right ) \, , 
\label{sconduc1}
\ee  
where 
\be 
\Gt_{\rm line} =  \frac{1}{4\pi (\sigma^2 -\tau^2)} = \frac{1}{4\pi} \, .  
\label{sline}
\ee 
We see that the spin conductance differs from the electric one. In particular,  
the conservation of the electric charge (\ref{kir2}) spoils the Kirchhoff's rule for $\Gt_{ij}$. 
Therefore, the spin is not a conserved quantum number in this case. Alternatively, 
one can impose the Kirchhoff's rule (\ref{kir4}), which guarantees spin conservation 
but breaks down the charge conservation.  

Comparing (\ref{conduc1}) and (\ref{sconduc1}), one discovers a simple but deep interplay 
between charge and spin transport: enhancement of the electric conductance corresponds 
to reduction of the spin conductance and vice versa. This feature clearly shows 
an effective separation in the dynamics of the charge and spin degrees of freedom of 
the model, extending to $\Gamma$ the charge-spin separation \cite{Voit} in the 
Luttinger model on the line.  

\section{Outlook and perspectives} 

Quantum field theory on graphs is an interesting subject both from the 
physical and the mathematical point of view. This contribution describes some 
fundamental aspects of this theory on a star graph $\Gamma$. First of all we discussed 
the impact of the Kirchhoff's rule (associated with any conserved current) on the
symmetry  content of the theory. In this context we established the crucial role of the
energy-momentum  tensor for selecting all boundary conditions which do not dissipate
energy in the  vertex of $\Gamma$. We described in detail the massless scalar field $\ph$ 
and its dual $\phd$, which represent the basic building blocks for the 
construction of vertex operators on $\Gamma$. 
Our construction is based on the point-like character of the boundary interactions 
at the vertex, the theory of self-adjoint extension of Hermitian operators 
on graphs \cite{Kostrykin:1998gz}-\cite{KSch} and the algebraic technique 
\cite{Liguori:1996xr}, \cite{Liguori:1997vd}, \cite{Mintchev:2001aq}-\cite{Mintchev:2007qt} for 
dealing with defects in QFT. 
In this framework we derived the Casimir energy density on $\Gamma$ and  
classified all scale invariant boundary conditions (critical points). 
We computed also the two-point correlation functions of the vertex operators 
and extracted their scaling dimensions at any critical point. Extending the 
method of bosonization on star graphs, 
we analyzed the four-fermion bulk interaction. Imposing 
scale invariant boundary conditions in terms of the electric current, we  
solved the massless four-fermion interaction exactly. This result was applied for  
the study of the charge and spin transport along $\Gamma$, which show interesting 
physical properties. 

The content of the above investigation can be generalized in several directions, 
which can certainly help a better understanding of the physics of quantum wires. 
Among others, we have in mind the study of generic graphs, integrable and more 
involved bulk interactions, finite temperature systems and dissipative phenomena 
at the vertex. 

\bigskip

\noindent{\bf Acknowledgments} 
\bigskip 

\noindent We thank Prof. Exner and Prof. Kuchment for 
the opportunity to present the above results at the workshop ``Graph models and 
mesoscopic systems, wave-guides and nano-structures". The first three authors 
would like to thank also the Isaac Newton Institute for Mathematical Sciences for the kind 
hospitality.

\end{document}